\documentclass[12pt]{iopart}
\usepackage{graphicx}

\begin{document}

\title[
Relating Quark Confinement and Chiral Symmetry Breaking in QCD]
{Relating Quark Confinement and Chiral Symmetry Breaking in QCD}
\author{Hideo Suganuma} 
\address{Department of Physics, Kyoto University, Kyoto 606-8502, Japan}
\author{Takahiro M. Doi} 
\address{Theoretical Research Division, Nishina Center, RIKEN, Saitama 351-0198, Japan}
\author{Krzysztof Redlich, Chihiro Sasaki} 
\address{Institute of Theoretical Physics, University of Wroclaw, 
PL-50204 Wroclaw, Poland}
\vspace{10pt}
\begin{indented}
\item[]Augst 2017
\end{indented}

\def\Slash#1{\not\!\!#1}

\begin{abstract}
We study the relation between quark confinement and 
chiral symmetry breaking in QCD.
Using lattice QCD formalism, we analytically express 
the various ``confinement indicators'', such as the Polyakov loop, its fluctuations, 
the Wilson loop, the inter-quark potential and the string tension, 
in terms of the Dirac eigenmodes. 
In the Dirac spectral representation, there appears 
a power of the Dirac eigenvalue $\lambda_n$ such as $\lambda_n^{N_t-1}$, 
which behaves as a reduction factor for small $\lambda_n$. 
Consequently, since this reduction factor cannot be cancelled, 
the low-lying Dirac eigenmodes give negligibly 
small contribution to the confinement quantities,
while they are essential for chiral symmetry breaking.
These relations indicate no direct, one-to-one correspondence 
between confinement and chiral symmetry breaking in QCD.
In other words, there is some independence of quark confinement 
from chiral symmetry breaking, which can generally lead to 
different transition temperatures/densities 
for deconfinement and chiral restoration. 
We also investigate the Polyakov loop in terms of the eigenmodes of 
the Wilson, the clover and the domain-wall fermion kernels, 
respectively, and find the similar results.
The independence of quark confinement 
from chiral symmetry breaking seems to be natural, 
because confinement is realized independently of quark masses 
and heavy quarks are also confined even without the chiral symmetry.
\end{abstract}

%
%
%
%
%

\section{Introduction}

About 50 years ago, Nambu first proposed an SU(3) gauge theory (QCD)
and introduced the SU(3) gauge field (gluons) for the strong interaction \cite{N66}, 
just after the introduction of new degrees of freedom of ``color'' in 1965 \cite{HN65}. 
Since the field-theoretical proof on the asymptotic freedom of QCD in 1973 \cite{GW73,P73}, 
QCD has been established as the fundamental theory of the strong interaction. 
In particular, perturbative QCD is quite successful for the description of 
high-energy hadron reactions in the framework of the parton model \cite{F69,BP69,GSS07}. 

In the low-energy region, however, QCD is a fairly difficult theory 
because of its strong-coupling nature, 
and shows nonperturbative properties such as color confinement and 
spontaneous chiral-symmetry breaking \cite{NJL61,BC80}. 
Here, spontaneous symmetry breaking appears in various physical phenomena, 
whereas the confinement is highly nontrivial and is never observed in most fields of physics.

For such nonperturbative phenomena, lattice QCD formalism 
was developed as a robust approach based on QCD \cite{W74,KS75}, and 
M.~Creutz first performed lattice QCD Monte Carlo calculations around 1980 \cite{C7980}.
Owing to lots of lattice QCD studies \cite{R12} so far, 
many nonperturbative aspects of QCD have been clarified to some extent. 
Nevertheless, most of nonperturbative QCD is not well understood still now. 

In particular, the relation between quark confinement and 
spontaneous chiral-symmetry breaking is not clear, 
and this issue has been an important difficult problem in QCD for a long time. 
A strong correlation between confinement and chiral symmetry breaking 
has been suggested by approximate coincidence between deconfinement and 
chiral-restoration temperatures \cite{R12,K02}, 
although it is not clear whether this coincidence is quantitatively precise or not.
At the physical point of the quark masses $m_{u,d}$ and $m_s$, 
a lattice QCD work \cite{AFKS06} shows about 25MeV difference between 
deconfinement and chiral-restoration temperatures, i.e., 
$T_{\rm deconf}\simeq 176 {\rm MeV}$ and $T_{\rm chiral}\simeq 151 {\rm MeV}$, 
whereas a recent lattice QCD study \cite{BBDPSVH16} shows 
almost the same transition temperatures of $T_{\rm deconf}\simeq T_{\rm chiral}$, 
after solving the scheme dependence of the Polyakov loop.

The correlation of quark confinement with chiral symmetry breaking 
is also suggested in terms of QCD-monopoles \cite{SST95,M95,W95}, 
which topologically appear in QCD in the maximally Abelian gauge. 
These two nonperturbative phenomena simultaneously lost by removing 
the QCD-monopoles from the QCD vacuum generated in lattice QCD \cite{M95,W95}.
This lattice QCD result surely indicates an important role of QCD-monopoles 
to confinement and chiral symmetry breaking, and they could have some relation
through the monopole, but their direct relation is still unclear.

As an interesting example, 
a recent lattice study of SU(2)-color QCD with $N_f=2$ 
exhibits that a confined but chiral-restored phase is realized 
at a large baryon density \cite{BIKMN16}. 
We also note that a large difference between confinement and 
chiral symmetry breaking actually appears 
in some QCD-like theories as follows:
\begin{itemize}
\item
In an SU(3) gauge theory with adjoint-color fermions,
the chiral transition occurs at much higher temperature than deconfinement, 
$T_{\rm chiral} \simeq 8~T_{\rm deconf}$ \cite{KL99}.
\item
In 1+1 QCD with $N_f \ge 2$, confinement is realized, 
whereas spontaneous chiral-symmetry breaking never occur, 
because of the Coleman-Mermin-Wagner theorem.
\item
In $N=1$ SUSY 1+3 QCD with $N_f=N_c+1$,
while confinement is realized, chiral symmetry breaking does not occur.
\end{itemize}
In any case, the relation between confinement and 
chiral symmetry breaking is an important open problem 
in wider theoretical studies including QCD, 
and many studies have been done so far
\cite{SST95,M95,W95,G06,SWL08,L11,GIS12,DSI14,DRSS15,SDI16,SDRS16}.

One of the key points to deal with chiral symmetry breaking 
is low-lying Dirac eigenmodes, 
since Banks and Casher discovered that these modes play the essential role 
to chiral symmetry breaking in 1980 \cite{BC80}. 
In this paper, keeping in mind the essential role of 
low-lying Dirac modes to chiral symmetry breaking, 
we derive {\it analytical relations} \cite{DSI14,DRSS15,SDI16,SDRS16} 
between the Dirac modes and the ``confinement indicators'', such as the Polyakov loop, 
the Polyakov-loop fluctuations \cite{LFKRS13a,LFKRS13b}, 
the Wilson loop, the inter-quark potential and the string tension, 
in lattice QCD formalism. 
Through the analyses of these relations, we investigate the correspondence 
between confinement and chiral symmetry breaking in QCD.

The organization of this paper is as follows. In Sec.~2, we briefly review 
the Dirac operator, and its eigenvalues and eigenmodes in lattice QCD.
In Sec.~3, we derive an analytical relation between the Polyakov loop 
and Dirac modes in temporally odd-number lattice QCD, 
and show independence of quark confinement from chiral symmetry breaking.
In Sec.~4, we investigate deconfinement and chiral restoration in finite temperature QCD, 
through the relation of the Polyakov-loop fluctuations with Dirac eigenmodes.
In Sec.~5, we derive analytical relations of the Polyakov loop 
with Wilson, clover and domain wall fermions, respectively. 
In Sec.~6, we express the Wilson loop with Dirac modes on arbitrary square lattices, 
and investigate the string tension, quark confining force, 
in terms of chiral symmetry breaking.
Section~7 is devoted to summary and conclusion.

Through this paper, we take {\it normal temporal (anti-)periodicity} 
for gauge and fermion variables, i.e., gluons and quarks, 
as is necessary to describe the thermal system properly, 
on an ordinary square lattice with the spacing $a$ 
and the size $V \equiv N_s^3 \times N_t$.
For the simple notation, we mainly use the lattice unit of $a=1$, 
although we explicitly write $a$ as necessary.
The numerical calculation is done in SU(3) quenched lattice QCD. 

\section{Dirac operator, Dirac eigenvalues and Dirac modes in lattice QCD}

In this section, we briefly review the Dirac operator $\hat{\Slash D}$, 
Dirac eigenvalues $\lambda_n$ and Dirac modes $|n \rangle$ 
in lattice QCD, where the gauge variable is described by 
the link-variable $U_\mu(s) \equiv {\rm e}^{iagA_\mu(s)} \in {\rm SU}(N_c)$ 
with the gauge coupling $g$ and the gluon field $A_\mu(x) \in {\rm su}(N_c)$. 

For the simple notation, we use 
$U_{-\mu}(s)\equiv U^\dagger_\mu(s-\hat \mu)$, 
and introduce the link-variable operator 
$\hat U_{\pm \mu}$ defined by the matrix element 
\cite{GIS12,DSI14,DRSS15,SDI16} 
\begin{eqnarray}
\langle s |\hat U_{\pm \mu}|s' \rangle 
=U_{\pm \mu}(s)\delta_{s\pm \hat \mu,s'}, 
\label{eq:LVO}
\end{eqnarray}
which satisfies $\hat U_{-\mu}=\hat U_\mu^\dagger$.
In lattice QCD formalism, 
the simple Dirac operator and the covariant derivative operator are given by 
\begin{eqnarray}
\hat{\Slash D}
=\frac{1}{2a}\sum_{\mu=1}^{4} \gamma_\mu (\hat U_\mu-\hat U_{-\mu}), \quad
\hat D_\mu
=\frac{1}{2a} (\hat U_\mu-\hat U_{-\mu}).
\label{eq:Dirac}
\end{eqnarray}
The Dirac operator $\hat{\Slash D}$ is anti-hermite satisfying 
\begin{eqnarray}
\hat{\Slash{D}}_{s',s}^\dagger=-\hat{\Slash{D}}_{s,s'},
\end{eqnarray}
so that its eigenvalue is pure imaginary.
We define the normalized Dirac eigenmode $|n \rangle$ 
and the Dirac eigenvalue $\lambda_n$, 
\begin{eqnarray}
\hat{\Slash{D}} |n\rangle =i\lambda_n |n \rangle \quad (\lambda_n \in {\bf R}), 
\qquad
\langle m|n\rangle=\delta_{mn}.
\end{eqnarray}
Since the eigenmode of any (anti)hermite operator generally makes a complete set, 
the Dirac eigenmode $|n \rangle$ of anti-hermite $\hat{\Slash D}$ 
satisfies the complete-set relation, 
\begin{eqnarray}
\sum_n |n \rangle \langle n|=1.
\label{eq:complete}
\end{eqnarray}
For the Dirac eigenfunction $\psi_n(s)\equiv\langle s|n \rangle$, 
the Dirac eigenvalue equation in lattice QCD is written as 
\begin{eqnarray}
\sum_{s'}\hat{\Slash D}_{s,s'} \psi_n(s')=i\lambda_n \psi_n(s),
\end{eqnarray}
and its explicit form is given by 
\begin{eqnarray}
\frac{1}{2a} \sum_{\mu=1}^4 \gamma_\mu
[U_\mu(s)\psi_n(s+\hat \mu)-U_{-\mu}(s)\psi_n(s-\hat \mu)] 
=i\lambda_n \psi_n(s).
\label{eq:Dirac-eigen-eq}
\end{eqnarray}

For the thermal system, considering the temporal anti-periodicity 
in $\hat D_4$ acting on quarks \cite{BBW13}, 
it is convenient to add a minus sign to the matrix element of 
the temporal link-variable operator $\hat U_{\pm 4}$ 
at the temporal boundary of $t=N_t$ as \cite{SDI16} 
\begin{eqnarray}
\langle {\bf s}, N_t|\hat U_4| {\bf s}, 1 \rangle 
&=&-U_4({\bf s}, N_t), \cr
\langle {\bf s}, 1|\hat U_{-4}| {\bf s}, N_t \rangle 
&=&-U_{-4}({\bf s}, 1)=-U_4^\dagger({\bf s}, N_t),
\label{eq:LVthermal}
\end{eqnarray}
which keeps $\hat U_{-\mu}=\hat U_\mu^\dagger$.
The standard Polyakov loop $L_P$ defined with $U_4(s)$ 
is simply written as the functional trace of 
$\hat U_4^{N_t}$,
\begin{eqnarray}
L_P \equiv \frac{1}{N_c V}
\sum_s {\rm tr}_c \{\prod_{n=0}^{N_t-1} U_4(s+n\hat t)\}
= -\frac{1}{N_c V} {\rm Tr}_c \{\hat U_4^{N_t}\},
\label{eq:PL}
\end{eqnarray}
with the four-dimensional lattice volume $V \equiv N_s^3 \times N_t$,
the functional trace ${\rm Tr}_c \equiv \sum_s {\rm tr}_c$, 
and the trace ${\rm tr}_c$ over color index.
The minus sign stems from the additional minus on $U_4({\bf s}, N_t)$, 
which reflects the temporal anti-periodicity of $\Slash D$ \cite{SDI16,BBW13}.

We here comment on the gauge ensemble average 
and the functional trace of operators in lattice QCD.
In the numerical calculation process of lattice QCD, 
one first generates many gauge configurations 
by importance sampling with the Monte Carlo method, 
and next evaluates the expectation value 
of the operator in consideration 
at each gauge configuration, 
and finally takes its gauge ensemble average.
In lattice QCD, the functional trace is expressed by 
a sum over all the space-time site, i.e., ${\rm Tr} = \sum_s {\rm tr}$, 
which is defined for each lattice gauge configuration. 
On enough large volume lattice, e.g., $N_s \rightarrow \infty$, 
the functional trace is proportional to the gauge ensemble average, 
${\rm Tr}~\hat O =\sum_s {\rm tr}~\hat O
 \propto \langle \hat O \rangle_{\rm gauge~ave.}$, 
for any operator $\hat O$.
Note also that, from the definition of $\hat U_{\pm \mu}$ in Eq.(\ref{eq:LVO}), 
the functional trace of any product of link-variable operators 
corresponding to ``{\it non-closed line}'' is {\it exactly zero}, i.e.,
\begin{eqnarray}
{\rm Tr}(\prod_{k=1}^N \hat U_{\mu_k})=0
\quad {\rm for} \quad \sum_{k=1}^N \mu_k \ne 0 
\qquad \mu_k \in \{\pm 1,\pm2,\pm3,\pm4\}, 
\label{eq:non-closed}
\end{eqnarray}
at each lattice gauge configuration \cite{DSI14,DRSS15,SDI16}, 
before taking the gauge ensemble average.

\begin{figure}[h]
\vspace{-1cm}
\begin{center}
\includegraphics[scale=0.5]{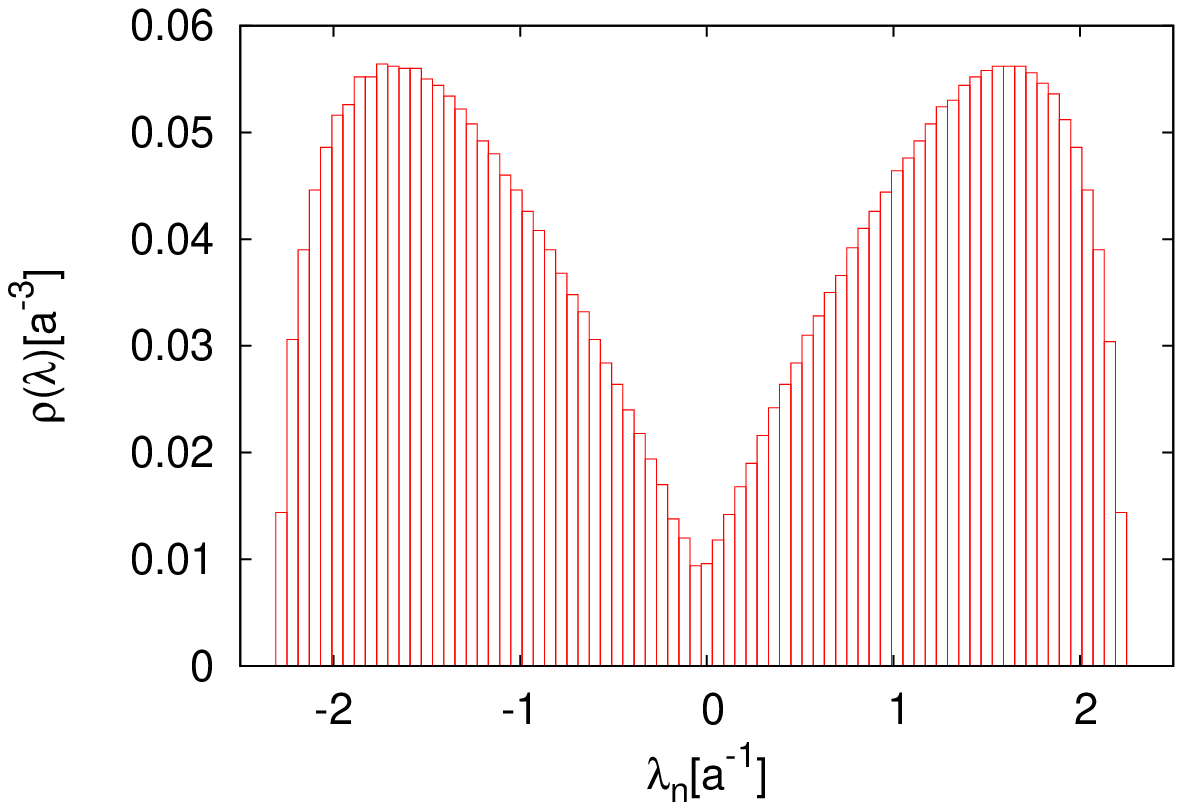}
\includegraphics[scale=0.5]{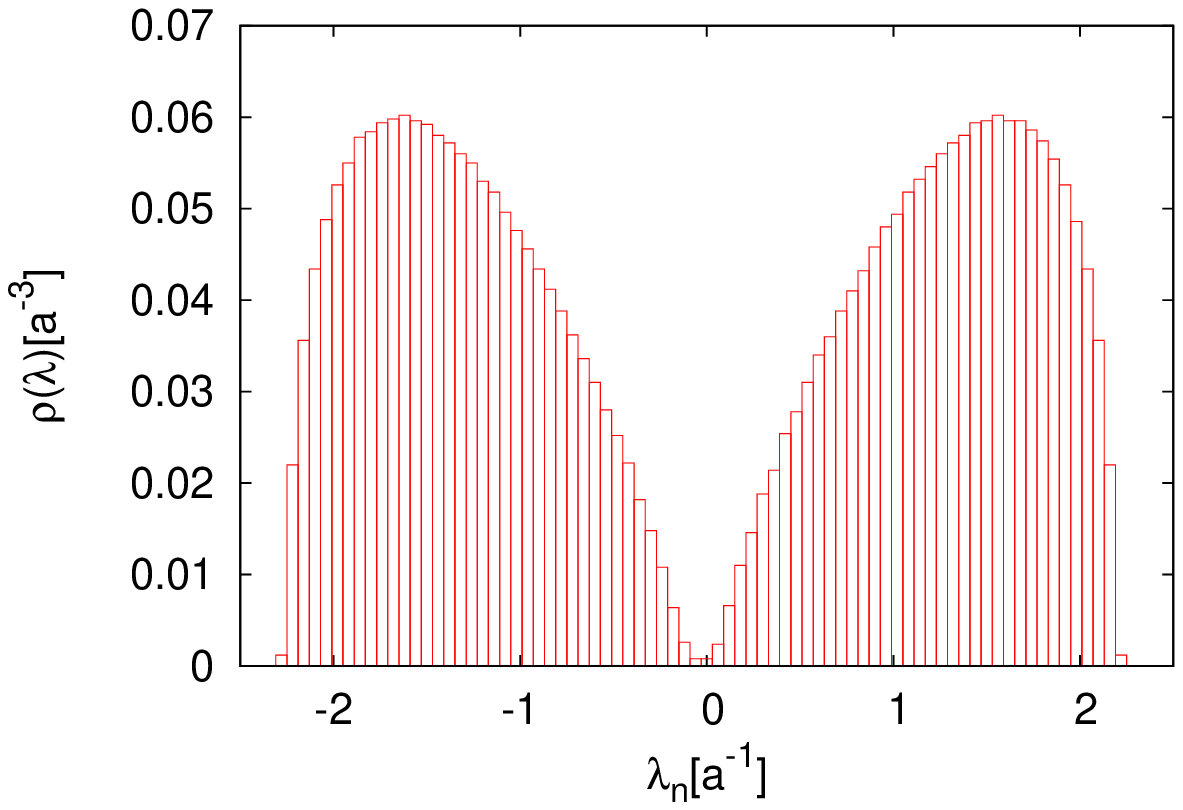}
\vspace{0.5cm}
\caption{
The lattice QCD result of the Dirac eigenvalue distribution 
$\rho(\lambda)$ in the lattice unit for 
(a) the confinement phase ($\beta = 5.6$, $10^3\times5$) and 
(b) the deconfinement phase ($\beta = 6.0$, $10^3\times5$). 
This figure is taken from Ref.~\cite{DRSS15}.
}
\label{fig:Dirac}
\end{center}
\vspace{-1cm}
\end{figure}

To finalize in this section, we briefly introduce the crucial role of 
low-lying Dirac modes to chiral symmetry breaking.
For each gauge configuration, 
the Dirac eigenvalue distribution is defined by
\begin{eqnarray}
\rho(\lambda) \equiv \frac{1}{V}\sum_n \delta(\lambda-\lambda_n),
\end{eqnarray} 
and then the Dirac zero-eigenvalue density $\rho(0)$ relates 
to the chiral condensate $\langle \bar qq \rangle$ via 
the Banks-Casher relation \cite{BC80}
\begin{eqnarray}
|\langle \bar qq \rangle|
= \lim_{m \rightarrow 0} \lim_{V \rightarrow \infty} \pi \rho(0).
\end{eqnarray}
%
The quantitative importance of $\rho(0)$ to chiral symmetry breaking is actually 
observed in lattice QCD simulations.
For example, Figs.~\ref{fig:Dirac}(a) and (b) show the lattice QCD result of 
the Dirac eigenvalue distribution $\rho(\lambda)$ 
in confined and deconfined phases, respectively. 
The near-zero Dirac-mode density 
$\rho(\lambda \simeq 0)$ takes a non-zero finite value in the confined phase, 
whereas it is almost zero in the deconfined phase.
In this way, the low-lying Dirac modes can be regarded as 
the essential modes for chiral symmetry breaking.

\section{Polyakov loop and Dirac modes in temporally odd-number lattice QCD}

In this section, we study the Polyakov loop and Dirac modes 
in temporally odd-number lattice QCD \cite{DSI14,DRSS15,SDI16}, 
where the temporal lattice size $N_t(< N_s)$ is odd. 
Note that, in the continuum limit of $a \rightarrow 0$ with 
keeping $N_t a$ constant, any large number $N_t$ gives the same physical result.
Then, it is no problem to use the odd-number lattice.

\begin{figure}[h]
\begin{center}
\includegraphics[scale=0.35]{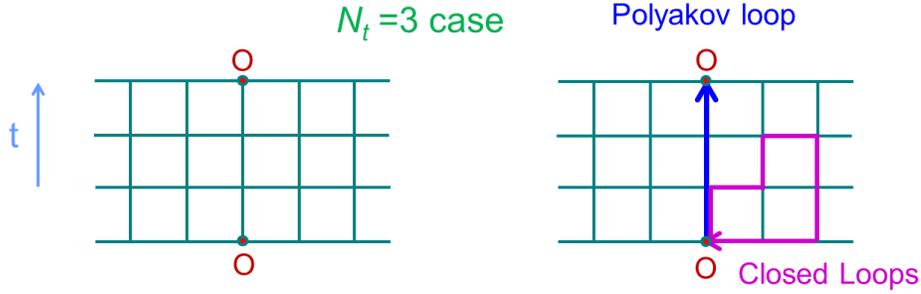}
\caption{
(a) A simple example of the temporally odd-number lattice ($N_t=3$ case).
(b) Only gauge-invariant quantities such as 
closed loops and the Polyakov loop survive in QCD.
Closed loops have even-number links on the square lattice.
}
\label{fig:loop}
\end{center}
\end{figure}

In general, only gauge-invariant quantities such as closed loops 
and the Polyakov loop survive in QCD, according to the Elitzur theorem \cite{R12}.
All the non-closed lines are gauge-variant and their expectation values are zero.
[Their functional traces are also exactly zero as Eq.(\ref{eq:non-closed}).] 
Note that any closed loop needs even-number link-variables 
on the square lattice, except for the Polyakov loop, as shown in Fig.\ref{fig:loop}.

\subsection{Analytical relation between Polyakov loop and Dirac modes}

\begin{figure}[h]
\begin{center}
\includegraphics[scale=0.24]{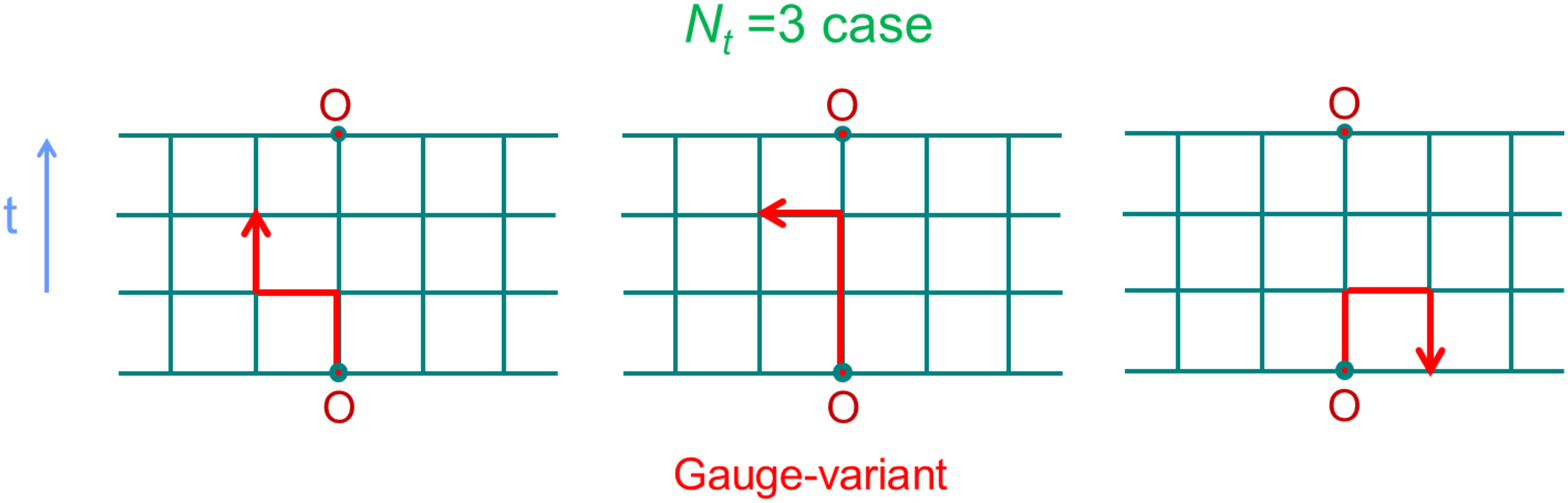}
\hspace{0.2cm}
\includegraphics[scale=0.24]{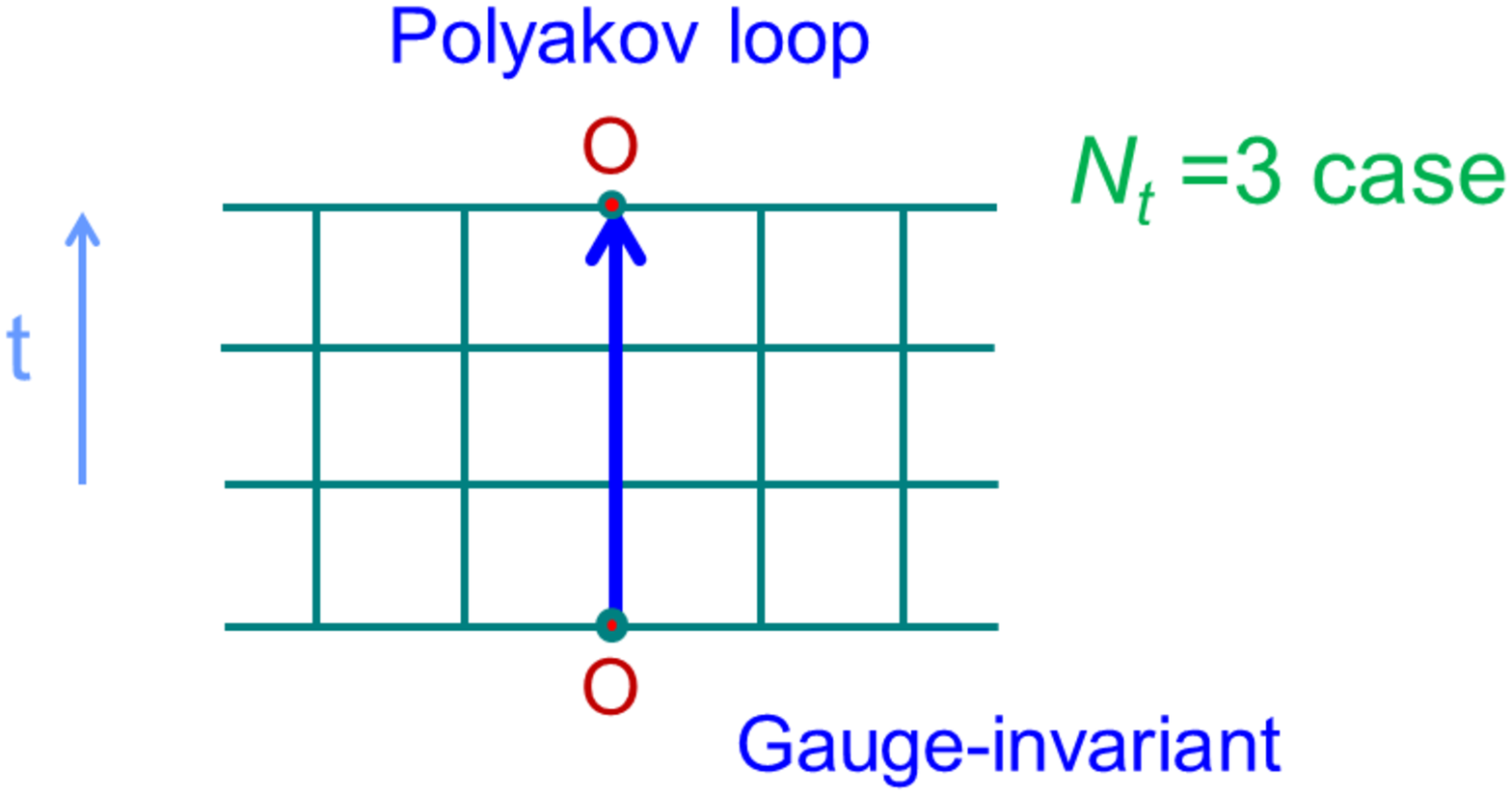}
\caption{
Examples of the trajectories stemming from 
$I ={\rm Tr}_{c,\gamma}(\hat U_4\hat{\Slash{D}}^{N_t-1})$. 
For each trajectory, the total length is 
$N_t$, and the ``first step'' is positive 
temporal direction corresponding to $\hat U_4$.
All the trajectories with the odd-number length $N_t$ 
cannot form a closed loop on the square lattice,
so that they are gauge-variant and give no contribution, 
except for the Polyakov loop.
}
\label{fig:line}
\end{center}
\end{figure}

On the temporally odd-number lattice, 
we consider the functional trace 
\cite{DSI14,DRSS15,SDI16}, 
\begin{eqnarray}
I\equiv {\rm Tr}_{c,\gamma} (\hat{U}_4\hat{\Slash{D}}^{N_t-1}), 
\label{eq:FTV}
\end{eqnarray}
where ${\rm Tr}_{c,\gamma}\equiv \sum_s {\rm tr}_c 
{\rm tr}_\gamma$ 
includes the sum over all the four-dimensional site $s$
and the traces over color and spinor indices. 
%
%
From Eq.(\ref{eq:Dirac}), $\hat U_4\hat{\Slash{D}}^{N_t-1}$ 
is written as a sum of products of $N_t$ link-variable operators, 
since the lattice Dirac operator $\hat{\Slash D}$ 
includes one link-variable operator $\hat U$ in each direction of $\pm \mu$.
Then, $\hat U_4\hat{\Slash{D}}^{N_t-1}$ is expressed as a sum of 
many Wilson lines with the total length $N_t$, as shown in Fig.\ref{fig:line}.

Note that all the trajectories with the odd-number length $N_t$ 
cannot form a closed loop on the square lattice, 
and corresponding Wilson lines are gauge-variant, except for the Polyakov loop.
In fact, almost all the trajectories stemming from 
$I = {\rm Tr}_{c,\gamma}(\hat U_4 \hat{\Slash D}^{N_t-1})$ 
is non-closed and give no contribution, whereas 
only the Polyakov-loop component in $I$ can survive 
as a gauge-invariant quantity. 
Thus, $I$ is proportional to the Polyakov loop $L_P$.
Actually, using Eqs.(\ref{eq:PL}) and (\ref{eq:non-closed}), 
one can mathematically derive the relation of 
\begin{eqnarray}
I &=& {\rm Tr}_{c,\gamma} (\hat U_4 \hat{\Slash D}^{N_t-1}) 
= {\rm Tr}_{c,\gamma} \{\hat U_4 (\gamma_4 \hat D_4)^{N_t-1}\} 
=4 {\rm Tr}_{c} (\hat U_4 \hat D_4^{N_t-1}) 
\cr
&=&\frac{4}{2^{N_t-1}} {\rm Tr}_{c} \{ \hat U_4 (\hat U_4-\hat U_{-4})^{N_t-1} \} 
=\frac{4}{2^{N_t-1}} {\rm Tr}_{c} \{ \hat U_4^{N_t} \} 
=-\frac{4N_cV}{2^{N_t-1}} L_P,
\label{eq:FTtoPL}
\end{eqnarray}
where the last minus reflects the temporal anti-periodicity of 
$\hat{\Slash{D}}$ \cite{SDI16,BBW13}.

On the other hand, using the completeness of the Dirac mode, 
$\sum_n |n\rangle \langle n|=1$, we calculate the functional trace 
in Eq.(\ref{eq:FTV}) and find the Dirac-mode representation of 
\begin{eqnarray}
I = \sum_n\langle n|\hat{U}_4 \hat{\Slash D}^{N_t-1}|n\rangle
=i^{N_t-1}\sum_n\lambda_n^{N_t-1}\langle n|\hat{U}_4| n \rangle. 
\label{eq:FTtoD}
\end{eqnarray}

Combing Eqs.(\ref{eq:FTtoPL}) and (\ref{eq:FTtoD}), 
we obtain the analytical relation between the Polyakov loop $L_P$ 
and the Dirac modes in QCD on the temporally odd-number lattice 
\cite{DSI14,DRSS15,SDI16}, 
\begin{eqnarray}
L_P=-\frac{(2i)^{N_t-1}}{4N_cV}
\sum_n\lambda_n^{N_t-1}\langle n|\hat{U}_4| n \rangle, 
\label{eq:PLvsD}
\end{eqnarray}
which is mathematically robust in both confined and deconfined phases. 
From Eq.(\ref{eq:PLvsD}), one can investigate each Dirac-mode contribution 
to the Polyakov loop individually.

Using the Dirac eigenvalue distribution 
$
\rho(\lambda)\equiv \frac{1}{V}\sum_n \delta(\lambda-\lambda_n),
$
this relation can be rewritten as
\begin{eqnarray}
L_P=-\frac{(2i)^{N_t-1}}{4N_c}
\int_{-\infty}^\infty d\lambda \rho(\lambda)
\lambda^{N_t-1} U_4(\lambda), 
\label{eq:PLvsDR}
\end{eqnarray}
with $U_4(\lambda_n) \equiv \langle n|\hat{U}_4| n \rangle$.
Here, the Dirac zero-eigenvalue density $\rho(0)$ relates to the chiral condensate 
$\langle \bar qq \rangle$ via the Banks-Casher relation, 
$|\langle \bar qq \rangle|
= \lim_{m \rightarrow 0} \lim_{V \rightarrow \infty} \pi \rho(0)$ \cite{BC80}.

As a remarkable fact, because of the strong reduction factor $\lambda_n^{N_t -1}$, 
low-lying Dirac-mode contribution 
is negligibly small in RHS of Eq.(\ref{eq:PLvsD}).
%
In Eq.(\ref{eq:PLvsDR}), the reduction factor $\lambda^{N_t-1}$ 
cannot be cancelled by other factors, 
because $\rho(0)$ is finite (not divergent) as the Banks-Casher relation suggests, 
and $U_4(\lambda)$ is 
also finite reflecting the compactness of $U_4 \in {\rm SU}(N_c)$.

To conclude, the low-lying Dirac modes give little contribution 
to the Polyakov loop, regardless of confined or deconfined phase \cite{DSI14,DRSS15,SDI16}.

\subsection{Properties on analytical relation between Polyakov loop and Dirac modes}

In order to emphasize the importance and generality of Eq.(\ref{eq:PLvsD}), 
we summarize its essential properties;
\begin{enumerate}
\item
The relation (\ref{eq:PLvsD}) is manifestly gauge invariant,
because of the gauge invariance of 
$
\langle n |\hat U_4|n\rangle =
\sum_s \langle n |s \rangle \langle s 
|\hat U_4| s+\hat t \rangle \langle s+\hat t|n\rangle
=\sum_s \psi_n^\dagger (s)U_4(s) \psi_n(s+\hat t)
$
under the gauge transformation, 
$\psi_n(s)\rightarrow V(s) \psi_n(s)$.
\item
In RHS, there is no cancellation between chiral-pair Dirac eigen-states, 
$|n \rangle$ and $\gamma_5|n \rangle$, since $(N_t-1)$ is even, i.e., 
$(-\lambda_n)^{N_t-1}=\lambda_n^{N_t-1}$, and 
$\langle n |\gamma_5 \hat U_4 \gamma_5|n\rangle
=\langle n |\hat U_4|n\rangle$. 
\item
The relation (\ref{eq:PLvsD}) is valid 
in a large class of gauge group of the theory, 
including SU($N_c$) gauge theory with any color number $N_c$.
\item
The relation (\ref{eq:PLvsD}) is also intact regardless of 
presence or absence of dynamical quarks, 
although dynamical-quark effects appear in $L_P$, 
the Dirac eigenvalue distribution $\rho(\lambda)$ and 
$\langle n |\hat U_4|n\rangle$. 
Equation~(\ref{eq:PLvsD}) remains valid at finite density and temperature. 
\end{enumerate}
Note here that Eq.(\ref{eq:PLvsD}) has the above-mentioned
generality and wide applicability, because 
our derivation is based on only a few setup conditions \cite{SDI16}:
\begin{itemize}
\item[1.] square lattice (including anisotropic cases) 
with temporal periodicity
\item[2.] odd-number temporal size $N_t (< N_s)$
\end{itemize}

In the following, we reconsider 
the technical merit to use the temporally odd-number lattice, 
i.e., absence of the loop contribution, by comparing with even-number lattices.
When $N_t$ is even, a similar argument can be applied through the relations 
\begin{eqnarray}
I' &\equiv& \Tr_{c,\gamma}(\gamma_4 \hat{U}_4 \hat {\Slash D}^{N_t-1})
=\sum_n\langle n|\gamma_4 \hat{U}_4 \hat{\Slash D}^{N_t-1}|n\rangle \cr
&=& i^{N_t-1}\sum_n\lambda_n^{N_t-1}\langle n|\gamma_4\hat{U}_4| n \rangle,
\end{eqnarray} 
however, apart from the Polyakov loop, 
there remains additional contribution from huge number of various loops 
(including reciprocating line-like trajectories) as
\begin{eqnarray}
I' &\equiv& \Tr_{c,\gamma}(\gamma_4 \hat{U}_4 \hat {\Slash D}^{N_t-1})
=\frac{1}{2^{N_t-1}} \Tr_{c,\gamma}[\gamma_4 \hat{U}_4 
\{\sum_{\mu=1}^4\gamma_\mu (\hat{U_\mu}-\hat{U}_{-\mu})\}^{N_t-1}] \cr
&=&-\frac{4N_cV}{2^{N_t-1}} L_P+({\rm loop~contribution}),
\end{eqnarray}
and thus one finds 
\begin{eqnarray}
L_P=-\frac{(2i)^{N_t-1}}{4N_cV}
\sum_n\lambda_n^{N_t-1}\langle n|\gamma_4 \hat{U}_4| n \rangle
+({\rm loop~contribution}),
\label{eq:even}
\end{eqnarray}
where the relation between the Polyakov loop $L_P$ and Dirac modes is not at all clear, 
owing to the complicated loop contribution.
In the continuum limit, both equations (\ref{eq:PLvsD}) and (\ref{eq:even}) simultaneously hold, and there appears some constraint on the loop contribution, 
which is not of interest.
In fact, by the use of the temporally odd-number lattice, 
all the additional loop contributions disappear, 
and one gets a transparent correspondence as Eq.(\ref{eq:PLvsD}) 
between the Polyakov loop and Dirac modes.
(For the other type of the formula on even-number lattices, 
see Appendix A in Ref.\cite{DSI14}.)

\subsection{Numerical analysis of the low-lying Dirac-mode contribution 
to the Polyakov loop}

Now, we perform lattice QCD Monte Carlo calculations 
on the temporally odd-number lattice, and investigate the relation (\ref{eq:PLvsD}) 
and the low-lying Dirac-mode contribution to the Polyakov loop.

To begin with, we calculate LHS and RHS in Eq.(\ref{eq:PLvsD}) independently, 
and compare these values in both confined and deconfined phases \cite{DSI14,DRSS15}.
The Polyakov loop $L_P$, i.e., the LHS, can be calculated 
in lattice QCD straightforwardly, following the definition in Eq.(\ref{eq:PL}).
For RHS in Eq.(\ref{eq:PLvsD}), 
we numerically solve the eigenvalue equation 
(\ref{eq:Dirac-eigen-eq}), and get all the eigenvalues $\lambda_n$ 
and the eigenfunctions $\psi_n(s)$. 
Once the Dirac eigenfunction $\psi_n(s)$ is obtained, 
the matrix element $\langle n |\hat U_4|n\rangle$ can be calculated as 
$
\langle n |\hat U_4|n\rangle =
\sum_s \langle n |s \rangle \langle s 
|\hat U_4| s+\hat t \rangle \langle s+\hat t|n\rangle
=\sum_s \psi_n^\dagger (s)U_4(s) \psi_n(s+\hat t)
$. 
In this way, RHS is evaluated in lattice QCD. 
(For more technical details, see Ref.\cite{DSI14}.)

As a numerical demonstration of Eq.(\ref{eq:PLvsD}), 
we calculate its LHS ($L_P$) and RHS (Dirac spectral sum) 
for each gauge configuration generated 
in quenched SU(3) lattice QCD,
and show representative values in confined and deconfined phases 
in Tables~\ref{table1} and \ref{table2}, respectively.
%
%
In both phases, 
one finds the exact relation $L_P={\rm RHS}$ in Eq.(\ref{eq:PLvsD}) 
at each gauge configuration. 
[Even in full QCD, the mathematical relation (\ref{eq:PLvsD}) 
must be valid, which is to be numerically confirmed in our future study.]

\begin{table}[h]
\caption{
Numerical lattice-QCD results for LHS ($L_P$) and RHS (Dirac spectral sum) 
of the relation (\ref{eq:PLvsD}) in the confinement phase 
with $\beta=5.6$ on $10^3\times5$ size lattice 
for each gauge configuration (labeled with ``Config. No.'') 
We also list $(L_P)_{\rm IR\hbox{-}cut}$ 
with the IR Dirac-mode cutoff of $\Lambda_{\rm IR}\simeq0.4 {\rm GeV}$.
This data is partially taken from Ref.~\cite{DSI14}.
}
\begin{tabular}{ccccccccccc} \hline \hline
Config. No. &1&2&3&4&5&6&7 \\ \hline
Re$ L_P $
&0.00961    &-0.00161     &0.0139    &-0.00324     &0.000689   
&0.00423    &-0.00807     \\
Re RHS
&0.00961    &-0.00161     &0.0139    &-0.00324     &0.000689   
&-0.00423   &-0.00807    \\
Re$(L_P)_{\rm IR\hbox{-}cut}$ 
&0.00961 &-0.00160&0.0139    &-0.00325&0.000706
&0.00422&-0.00807         \\ 
Im$ L_P $
&-0.00322  &-0.00125     &-0.00438 &-0.00519     &-0.0101       
&-0.0168    &-0.00265     \\
Im RHS
&-0.00322  &-0.00125     &-0.00438 &-0.00519     &-0.0101     
&-0.0168    &-0.00265     \\ 
Im$(L_P)_{\rm IR\hbox{-}cut}$ 
&-0.00321&-0.00125&-0.00437&-0.00520&-0.0101  
&-0.0168&-0.00264        \\ \hline \hline 
\end{tabular}
\label{table1}
\end{table}

\begin{table}[h]
\caption{
Numerical lattice-QCD results for LHS ($L_P$) and RHS (Dirac spectral sum) 
of the relation (\ref{eq:PLvsD}) in the deconfinement phase 
with $\beta=5.7$ on $10^3\times3$ size lattice 
for each gauge configuration (labeled with ``Config. No.'') 
We also list $(L_P)_{\rm IR\hbox{-}cut}$ 
with the IR Dirac-mode cutoff of $\Lambda_{\rm IR}\simeq0.4 {\rm GeV}$.
This data is partially taken from Ref.~\cite{DSI14}.
}
\begin{tabular}{ccccccccccc} \hline \hline
Config. No. &1&2&3&4&5&6&7 \\ \hline
Re$ L_P $
&0.316       &0.337       &0.331     &0.305      &0.313     
&0.316       &0.337       \\
Re RHS
&0.316       &0.337       &0.331     &0.305      &0.314     
&0.316       &0.337       \\
Re$(L_P)_{\rm IR\hbox{-}cut}$ 
&0.319     &0.340       &0.334     &0.307     &0.317   
&0.319      &0.340         \\
Im$ L_P $
&-0.00104  &-0.00597  &0.00723  &-0.00334 &0.00167  
&0.000120  &0.000482  \\
Im RHS
&-0.00104  &-0.00597   &0.00723  &-0.00334 &0.00167 
&0.000120  &0.000482  \\ 
Im$(L_P)_{\rm IR\hbox{-}cut}$ 
&-0.00103&-0.00597  &0.00724  &-0.00333&0.00167
&0.000121 &0.0000475 \\ \hline \hline
\end{tabular}
\label{table2}
\end{table}

Next, we introduce the IR cutoff of $\Lambda_{\rm IR} \simeq0.4 {\rm GeV}$ 
for the Dirac modes, and remove the low-lying Dirac-mode contribution 
for $|\lambda_n|<\Lambda_{\rm IR}$ from RHS in Eq.(\ref{eq:PLvsD}).
The chiral condensate after the removal of 
IR Dirac-modes below $\Lambda_{\rm IR}$ is expressed as 
\begin{eqnarray}
\langle \bar{q}q\rangle_{\Lambda_{\rm IR}}
=-\frac{1}{V}\sum_{|\lambda_n| \geq\Lambda_{\rm IR}}\frac{2m}{\lambda_n^2+m_q^2}.
\label{eq:chiral}
\end{eqnarray}
In the confined phase, this IR Dirac-mode cut leads to 
$\frac{\langle \bar{q}q\rangle_{\Lambda_{\rm IR}}}
{\langle \bar{q}q\rangle} \simeq 0.02$ and almost chiral-symmetry restoration 
in the case of physical current-quark mass, $m_q \simeq 5 {\rm MeV}$ \cite{DSI14}. 

We calculate $(L_P)_{\rm IR\hbox{-}cut}$ 
with the IR Dirac-mode cutoff of $\Lambda_{\rm IR}\simeq0.4 {\rm GeV}$, 
and add in Table \ref{table1} and Table \ref{table2} 
the results in both confined and deconfined phases, respectively. 
One finds that $L_P\simeq(L_P)_{\rm IR\hbox{-}cut}$ 
is satisfied for each gauge configuration in both phases. 
For the configuration average, 
$\langle L_P\rangle\simeq\langle (L_P)_{\rm IR\hbox{-}cut}\rangle$ 
is of course satisfied. 

From the analytical relation (\ref{eq:PLvsD}) and the numerical lattice QCD calculation, 
we conclude that low-lying Dirac-modes 
give negligibly small contribution to the Polyakov loop $L_P$, 
and are not essential for confinement, 
while these modes are essential for chiral symmetry breaking. 
This indicates no direct one-to-one correspondence between 
confinement and chiral symmetry breaking in QCD.

\section{Role of Low-lying Dirac modes to Polyakov-loop fluctuations}

\begin{figure}[h]
\begin{center}
\includegraphics[scale=0.5]{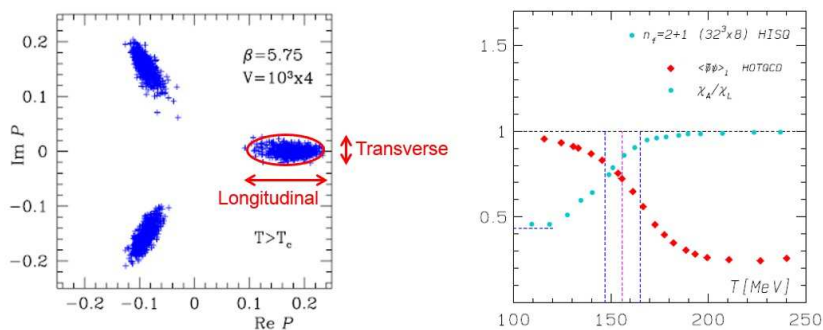}
\caption{
(a) The scatter plot of the Polyakov loop in lattice QCD. 
The original figure is taken from Ref.\cite{GHRSSW01}.
(b) The temperature dependence of the Polyakov loop susceptibilities ratio 
$R_A \equiv \chi_A/\chi_L$ and the light-quark chiral condensate 
$\langle \bar \psi \psi\rangle_l$, normalized to its zero temperature value 
on a finite lattice. 
The lattice QCD Monte Carlo data are from Refs. \cite{LFKRS13b} and \cite{HQCD14}, respectively. This figure is taken from Ref.\cite{DRSS15}.
}
\label{fig:PL}
\end{center}
\end{figure}

In thermal QCD, the Polyakov loop has fluctuations 
in longitudinal and transverse directions, as shown in Fig.\ref{fig:PL}(a), 
and the Polyakov-loop fluctuation gives a possible indicator of 
deconfinement transition \cite{LFKRS13a,LFKRS13b}. 
In this section, we investigate the Polyakov-loop fluctuations 
and the role of low-lying Dirac modes \cite{DRSS15} 
in the deconfinement transition at finite temperatures.

For the Polyakov loop $L_P$, 
we define its longitudinal and transverse components, 
\begin{eqnarray}
L_L \equiv {\rm Re}~\tilde L_P, \qquad L_T \equiv {\rm Im}~\tilde L_P,
\end{eqnarray}
with $\tilde L_P \equiv L_P~e^{2\pi i k/3}$ where $k \in \{0, \pm 1\}$ 
is chosen such that the $Z_3$-transformed Polyakov loop lies 
in its real sector \cite{DRSS15,LFKRS13a,LFKRS13b}.
We introduce the Polyakov-loop fluctuations as 
\begin{eqnarray}
\chi_A \propto \langle |L_P|^2 \rangle-|\langle L_P \rangle|^2, \quad
\chi_L \propto \langle L_L^2 \rangle-\langle L_L \rangle^2, \quad
\chi_T \propto \langle L_T^2 \rangle-\langle L_T \rangle^2,
\end{eqnarray}
and consider their different ratios which are known to largely change 
around the transition temperature \cite{LFKRS13a,LFKRS13b}, 
thus can be used as a good indicator of the deconfinement transition.

As an illustration, we show in Fig.\ref{fig:PL}(b) 
the temperature dependence of the ratio $R_A \equiv \chi_A/\chi_L$. 
Around the transition temperature, one finds a large change of $R_A$ 
on the Polyakov-loop fluctuation, which reflects the onset of deconfinement, 
while the rapid reduction of the chiral condensate indicates 
a signal of chiral symmetry restoration. 
As a merit to use the ratio $R_A$ instead of the individual fluctuations, 
multiplicative-type uncertainties related to the renormalization of 
the Polyakov loop are expected to be cancelled to a large extent in the ratio. 

Similarly in Sec.~3, 
we derive Dirac-mode expansion formulae for Polyakov-loop fluctuations
in temporally odd-number lattice QCD \cite{DRSS15}.
For the ratios $R_A \equiv \chi_A/\chi_L$ and $R_T \equiv \chi_T/\chi_L$, 
we obtain their Dirac spectral representation as 
\begin{eqnarray}
R_A=
\frac{
\left\langle\left|\sum_n \lambda_n^{N_t-1} \hat{U}_4^{nn} \right|^2\right\rangle
-
\left\langle\left|\sum_n \lambda_n^{N_t-1} \hat{U}_4^{nn}\right|\right\rangle^2
}{
\left\langle\left(\sum_n \lambda_n^{N_t-1}{\rm Re}\left(\mathrm{e}^{2\pi ki/3}\hat{U}_4^{nn}\right)\right)^2\right\rangle
-
\left\langle\sum_n \lambda_n^{N_t-1}{\rm Re}\left(\mathrm{e}^{2\pi ki/3}\hat{U}_4^{nn}\right)\right\rangle^2
}, 
\label{eq:PFA}
\end{eqnarray}
\begin{eqnarray}
R_T=
\frac{
\left\langle\left(\sum_n \lambda_n^{N_t-1}{\rm Im}\left(\mathrm{e}^{2\pi ki/3}\hat{U}_4^{nn}\right)\right)^2\right\rangle
-
\left\langle\sum_n \lambda_n^{N_t-1}{\rm Im}\left(\mathrm{e}^{2\pi ki/3}\hat{U}_4^{nn}\right)\right\rangle^2
}{
\left\langle\left(\sum_n \lambda_n^{N_t-1}{\rm Re}\left(\mathrm{e}^{2\pi ki/3}\hat{U}_4^{nn}\right)\right)^2\right\rangle
-
\left\langle\sum_n \lambda_n^{N_t-1}{\rm Re}\left(\mathrm{e}^{2\pi ki/3}\hat{U}_4^{nn}\right)\right\rangle^2
}, 
\label{eq:PFT}
\end{eqnarray}
with $\hat{U}_4^{nn} \equiv \langle n|\hat{U}_4| n \rangle$. 
We note that all the Polyakov-loop fluctuations are almost unchanged 
by removing low-lying Dirac modes \cite{DRSS15} 
in both confined and deconfined phases, because of 
the significant reduction factor $\lambda_n^{N_t-1}$ appearing in the Dirac-mode sum.

\begin{figure}[h]
\begin{center}
\includegraphics[scale=0.6]{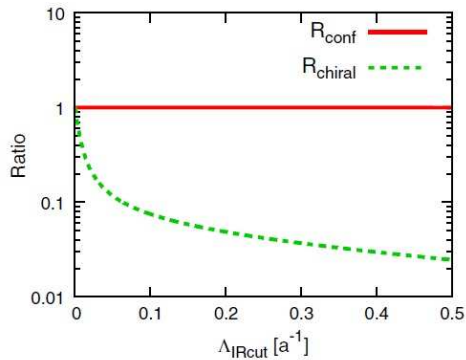}
\caption{
The lattice QCD result of IR Dirac-mode-removed quantities of 
$R_{\rm conf}(\Lambda_{\rm IRcut}) \equiv R_A(\Lambda_{\rm IRcut})/R_A$ 
and $R_{\rm chiral}(\Lambda_{\rm IRcut}) 
\equiv \langle \bar qq \rangle_{\Lambda_{\rm IRcut}}/\langle \bar qq \rangle$ 
for $m_q$= 5 MeV, 
plotted against the infrared cutoff $\Lambda_{\rm IRcut}$ introduced on 
Dirac eigenvalues, in quenched lattice with 
$\beta=5.6$ and $10^3 \times 5$ (confined phase).
This figure is taken from Ref.\cite{DRSS15}.
}
\label{fig:PF}
\end{center}
\end{figure}

Next, let us consider the actual removal of low-lying Dirac-mode contribution 
from the Polyakov-loop fluctuation using Eqs.(\ref{eq:PFA}) and (\ref{eq:PFT}), 
and also the chiral condensate for comparison. 
As an example, we show in Fig.\ref{fig:PF} the lattice QCD result of 
\begin{eqnarray}
R_{\rm conf}(\Lambda_{\rm IRcut}) \equiv \frac{R_A(\Lambda_{\rm IRcut})}{R_A}, \quad
R_{\rm chiral}(\Lambda_{\rm IRcut}) \equiv 
\frac{\langle \bar qq \rangle_{\Lambda_{\rm IRcut}}}{\langle \bar qq \rangle}
\end{eqnarray}
in the presence of the infrared Dirac-mode cutoff $\Lambda_{\rm IRcut}$ 
in the confined phase ($\beta=5.6$, $10^3 \times 5$) \cite{DRSS15}.
Here, $R_A(\Lambda_{\rm IRcut})$ denotes the truncated value of $R_A$ 
when the low-lying Dirac-mode contribution of $|\lambda_n|<\Lambda_{\rm IRcut}$ 
is removed from the Dirac spectral sum in Eq.(\ref{eq:PFA}). 
The truncated chiral condensate $\langle \bar qq \rangle_{\Lambda_{\rm IRcut}}$ 
is defined with Eq.(\ref{eq:chiral}), and the current-quark mass is taken as $m_q$=5MeV.

In contrast to the strong sensitivity of the chiral condensate 
$R_{\rm chiral}(\Lambda_{\rm IRcut})$, 
the Polyakov-loop fluctuation ratio $R_{\rm conf}(\Lambda_{\rm IRcut})$ 
is almost unchanged 
against the infrared cutoff $\Lambda_{\rm IRcut}$ of the Dirac mode \cite{DRSS15}, 
as discussed after Eq.(\ref{eq:PFT}).
Thus, we find no significant role of low-lying Dirac modes 
for the Polyakov-loop fluctuation as a new-type confinement/deconfinement indicator, 
which also indicates independence of confinement from chiral property in thermal QCD. 

\section{Relations of Polyakov loop with Wilson, clover and domain wall fermions}

All the above formulae are mathematically correct, 
because we have just used the Elitzur theorem 
(or precisely Eq.(\ref{eq:LVO}) for $\hat U_{\pm \mu}$) 
and the completeness $\sum_n|n \rangle \langle n|=1$ on the eigenmode 
$|n \rangle$ of the Dirac operator $\hat{\Slash D}$ in Eq.(\ref{eq:Dirac}).
Note here that we have not introduced dynamical fermions corresponding to 
the simple lattice Dirac operator $\hat{\Slash D}$, but only use 
the mathematical eigenfunction of $\hat{\Slash D}$.

However, one may wonder the doublers \cite{R12} in the use of 
the lattice Dirac operator $\hat{\Slash D}$ in Eq.(\ref{eq:Dirac}), 
although it is not the problem in the above formulae.
In fact, according to the Nielsen-Ninomiya theorem \cite{NielsenNinomiya}, 
$2^d$ modes simultaneously appear per fermion on a lattice,
if one uses a bilinear fermion action satisfying 
translational invariance, chiral symmetry, hermiticity and locality, 
on a $d$-dimensional lattice.
Then, for the description of hadrons without the redundant doublers, 
one adopts Wilson, staggered, clover, domain-wall (DW) or overlap fermion 
\cite{R12,SW85,NNMS03,K92,S93FS95,N98}, 
and has to deal with the fermion kernel $K$, 
which is more complicated than the simple Dirac operator $\hat{\Slash D}$.
In fact, the realization of a chiral fermion on a lattice is not unique. 
However, in the continuum limit, they must lead to the same physical result. 
Consequently, it is meaningful to examine our formulation 
with various fermion actions to get the robust physical conclusion.

In this section, we express the Polyakov loop with the eigenmodes of 
the kernel of the Wilson fermion, the clover ($O(a)$-improved Wilson) fermion 
and the DW fermion \cite{SDRS16}.
(The formulation with the overlap fermion is found in Ref.~\cite{D17}.)

\subsection{The Wilson fermion}

A simple way to remove the redundant fermion doublers 
is to make their mass extremely large by an additional interaction.
The Wilson fermion is constructed on a four-dimensional lattice 
by adding the $O(a)$ Wilson term \cite{R12}, 
which explicitly breaks the chiral symmetry.
For the Wilson fermion, all the doublers acquire a large mass of $O(a^{-1})$ 
and are decoupled at low energies near the continuum, $a \simeq 0$.
Thus, the Wilson fermion can describe the single light fermion, 
although it gives $O(a)$ explicit chiral-symmetry breaking on the lattice. 

The Wilson fermion kernel is described with 
the link-variable operator $\hat U_{\pm \mu}$ as \cite{SDRS16}
\begin{eqnarray}
\hat K&=&\hat {\Slash D}+m+
\frac{r}{2a}\sum_{\mu=\pm 1}^{\pm 4} \gamma_\mu (\hat U_\mu-1)
\cr
&=&\frac{1}{2a}\sum_{\mu=1}^{4} \gamma_\mu (\hat U_\mu-\hat U_{-\mu})
+m+\frac{r}{2a}\sum_{\mu=1}^4 \gamma_\mu (\hat U_\mu+\hat U_{-\mu}-2),
\label{eq:WFK}
\end{eqnarray}
which goes to $\hat K\simeq (\hat {\Slash D}+m)+ar \hat D^2$ 
near the continuum, $a \simeq 0$.
Thus, the Wilson term $ar \hat D^2$ is $O(a)$. The Wilson parameter $r$ is real.
Note that each term of $\hat K$ includes one $\hat U_{\pm \mu}$ at most, 
and connects only the neighboring site or acts on the same site.

For the Wilson fermion kernel $\hat K$, 
we define its eigenmode $|n \rangle \rangle$ and 
eigenvalue $\tilde \lambda_n$,
\begin{eqnarray}
\hat K| n \rangle \rangle =i \tilde \lambda_n |n \rangle \rangle, \quad 
\tilde \lambda_n \in {\bf C}.
\end{eqnarray}
If the Wilson term is absent, the eigenmode of $\hat K=\hat {\Slash D}+m$ is given by 
the simple Dirac eigenmode $|n\rangle$, i.e., $\hat K|n \rangle=(i\lambda_n+m)|n \rangle$, 
and satisfies the completeness of $\sum_n |n \rangle \langle n|=1$. 
In the presence of the $O(a)$ Wilson term, $\hat K$ is neither hermite nor anti-hermite, 
and the completeness generally includes an $O(a)$ error,
\begin{eqnarray}
\sum_n |n \rangle \rangle \langle \langle n|=1+O(a).
\label{eq:QCR}
\end{eqnarray}

\begin{figure}[h]
\begin{center}
\includegraphics[scale=0.5]{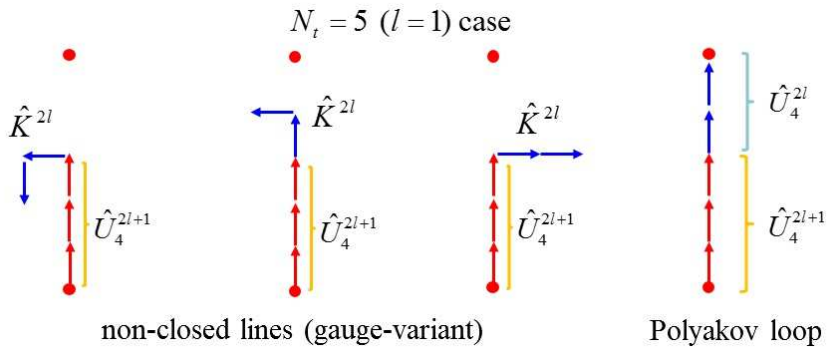}
\caption{
Some examples of the trajectories (corresponding to products of $\hat U$) 
in $J \equiv {\rm Tr}(\hat U_4^{2l+1}\hat K^{2l})$ for the $N_t=5$ ($l=1$) case. 
The length does not exceed $N_t$ for each trajectory. 
Only the Polyakov loop $L_P$ can form a closed loop and survives in $J$. 
}
\label{fig:FK}
\end{center}
\end{figure}

Here, we consider the functional trace $J$ on a lattice with $N_t =4l+1$ ($l=1,2,...$),
\begin{eqnarray}
J \equiv {\rm Tr}(\hat U_4^{2l+1} \hat K^{2l}).
\end{eqnarray}
By the use of the quasi-completeness (\ref{eq:QCR}) for $|n \rangle\rangle$, 
one finds, apart from an $O(a)$ error, 
\begin{eqnarray}
J \simeq \sum_n\langle \langle n|\hat{U}_4^{2l+1}{\hat{K}}^{2l}|n \rangle \rangle
=\sum_n(i \tilde \lambda_n)^{2l} \langle \langle n|\hat{U}_4^{2l+1}| n \rangle \rangle.
\end{eqnarray}
Since the kernel $\hat K$ in Eq.~(\ref{eq:WFK}) is a first-order equation of 
the link-variable operator $\hat U$, 
$J \equiv {\rm Tr}(\hat U_4^{2l+1} \hat K^{2l})$ is expressed as a sum of 
products of $\hat U$ with some $c$-number factor.
In each product in $J$, the total number of $\hat U$ does not exceed $N_t=4l+1$, 
because $\hat K$ is a first-order equation of $\hat U$.
Each product gives a trajectory as shown in Fig.\ref{fig:FK}.

\noindent
Among these many trajectories, 
only the Polyakov loop $L_P$ can form a closed loop 
and survives in $J$, which leads to $J \propto L_P$. 
Thus, apart from an $O(a)$ error, we obtain \cite{SDRS16}
\begin{eqnarray}
L_P \propto
\sum_n \tilde \lambda_n^{2l}
\langle \langle n|\hat{U}_4^{2l+1}| n \rangle \rangle.
\end{eqnarray}
Using the eigenvalue distribution 
$\rho(\tilde \lambda)\equiv \frac{1}{V}\sum_n \delta(\tilde \lambda-\tilde \lambda_n)$,
this relation is rewritten as
\begin{eqnarray}
L_P \propto
\int_{-\infty}^\infty d\tilde \lambda \rho(\tilde \lambda)
\tilde \lambda^{2l} u(\tilde \lambda), 
\end{eqnarray}
with $u(\tilde \lambda_n) \equiv 
\langle \langle n|\hat{U}_4^{2l+1}| n \rangle \rangle$.
As in Sec.3, the reduction factor $\tilde \lambda^{2l}$ 
cannot be cancelled by other factors, 
because of the finiteness of $\rho(0)$ and $u(\tilde \lambda)$ reflecting 
the Banks-Casher relation and the compactness of $U_4 \in {\rm SU}(N_c)$.
Thus, owing to the strong reduction factor $\tilde \lambda_n^{2l}$ 
on the eigenvalue $\tilde \lambda_n$ of the Wilson fermion kernel $\hat K$, 
we find also small contribution from low-lying modes of $\hat K$ to $L_P$.

\subsection{The clover (O(a)-improved Wilson) fermion}

The clover fermion is an $O(a)$-improved Wilson fermion \cite{SW85} 
with a reduced lattice discretization error of $O(a^2)$ near the continuum, 
and gives accurate lattice results \cite{NNMS03}. 
The clover fermion kernel is expressed as \cite{SDRS16}
\begin{eqnarray}
\hat K&=&\hat {\Slash D}+m
+\frac{r}{2a}\sum_{\mu=\pm 1}^{\pm 4} \gamma_\mu (\hat U_\mu-1)
+\frac{arg}{2}\sigma_{\mu\nu}G_{\mu\nu}, \cr
&=&\frac{1}{2a}\sum_{\mu=1}^{4} \gamma_\mu (\hat U_\mu-\hat U_{-\mu})
+m
+\frac{r}{2a}\sum_{\mu=1}^4 \gamma_\mu (\hat U_\mu+\hat U_{-\mu}-2) \cr
&+&\frac{arg}{2}\sigma_{\mu\nu}G_{\mu\nu},
\label{eq:CFK}
\end{eqnarray}
with $\sigma_{\mu\nu}\equiv \frac{i}{2}[\gamma_\mu, \gamma_\nu]$ 
and $G_{\mu\nu}$ being the clover-type lattice field strength defined by 
\begin{eqnarray}
G_{\mu\nu} &\equiv& \frac{1}{8}(P_{\mu\nu}+P^\dagger_{\mu\nu}),
\\ 
P_{\mu\nu}(x) &\equiv& 
\langle x|
(\hat U_\mu \hat U_\nu \hat U_{-\mu} \hat U_{-\nu}
+ \hat U_\nu \hat U_{-\mu} \hat U_{-\nu} \hat U_\mu
+ \hat U_{-\mu} \hat U_{-\nu} \hat U_\mu \hat U_\nu \cr
&+& \hat U_{-\nu} \hat U_\mu \hat U_\nu \hat U_{-\mu} )
|x \rangle.
\end{eqnarray} 
Since $G_{\mu\nu}$ acts on the same site, 
each term of $\hat K$ in Eq.(\ref{eq:CFK}) 
connects only the neighboring site or acts on the same site. 
Then, the technique done for Wilson fermions is also useful. 
 
For the clover fermion kernel $\hat K$, 
we define its eigenmode $|n \rangle \rangle$ and 
eigenvalue $\tilde \lambda_n$ as 
\begin{eqnarray}
\hat K| n \rangle \rangle =i \tilde \lambda_n |n \rangle \rangle, \quad 
\tilde \lambda_n \in {\bf C}, \quad \qquad
\sum_n |n \rangle \rangle \langle \langle n|=1+O(a^2).
\label{eq:QCR2}
\end{eqnarray}
Again, we consider the functional trace on a lattice with $N_t =4l+1$, 
\begin{eqnarray}
J \equiv {\rm Tr}(\hat U_4^{2l+1} \hat K^{2l}) 
\simeq \sum_n\langle \langle n|\hat{U}_4^{2l+1}{\hat{K}}^{2l}|n \rangle \rangle
=\sum_n(i \tilde \lambda_n)^{2l} \langle \langle n|\hat{U}_4^{2l+1}| n \rangle \rangle,
\end{eqnarray}
where we have used the quasi-completeness 
for $|n\rangle \rangle$ 
in Eq.(\ref{eq:QCR2}) within an $O(a^2)$ error. 
$J \equiv {\rm Tr}(\hat U_4^{2l+1} \hat K^{2l})$ is expressed as a sum of 
products of $\hat U$ with the other factor, 
and each product gives a trajectory as shown in Fig.\ref{fig:FK}.
Among the trajectories, 
only the Polyakov loop $L_P$ can form a closed loop 
and survives in $J$, i.e., $J \propto L_P$. 
Thus, apart from an $O(a^2)$ error, we obtain \cite{SDRS16}
\begin{eqnarray}
L_P \propto
\sum_n \tilde \lambda_n^{2l}
\langle \langle n|\hat{U}_4^{2l+1}| n \rangle \rangle, \qquad
L_P \propto
\int_{-\infty}^\infty d\tilde \lambda \rho(\tilde \lambda)
\tilde \lambda^{2l} u(\tilde \lambda), 
\end{eqnarray}
with $u(\tilde \lambda_n) \equiv 
\langle \langle n|\hat{U}_4^{2l+1}| n \rangle \rangle$ and 
$\rho(\tilde \lambda)\equiv \frac{1}{V}\sum_n \delta(\tilde \lambda-\tilde \lambda_n)$.
We thus find small contribution from low-lying modes of $\hat K$ 
to $L_P$, because of the suppression factor $\tilde \lambda_n^{2l}$ in the sum.

\subsection{The domain wall (DW) fermion}

\begin{figure}[h]
\begin{center}
\includegraphics[scale=0.4]{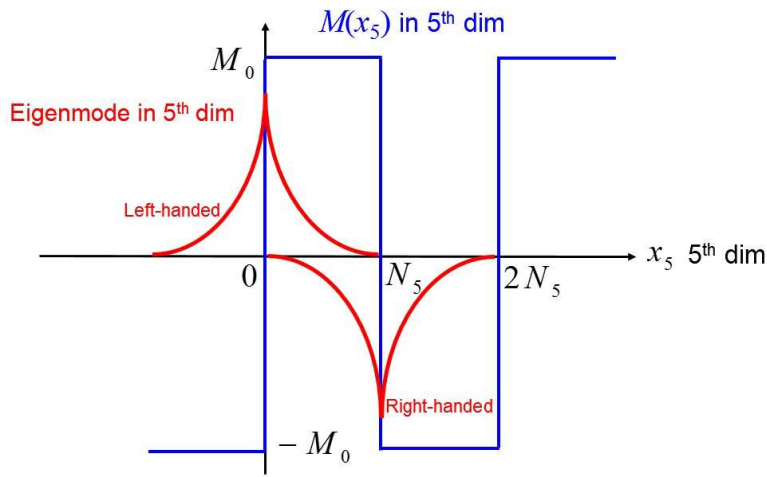}
\caption{
The construction of the domain wall (DW) fermion 
by introducing the fifth dimension of $x_5$ 
and the $x_5$-dependent mass $M(x_5)$. 
There appear left- and right-handed chiral zero modes localized 
around $x_5=0$ and $x_5=N_5$, respectively. 
}
\label{fig:DWF}
\end{center}
\end{figure}

Next, we consider the domain-wall (DW) fermion \cite{K92,S93FS95}, 
which realizes the ``exact'' chiral symmetry on a lattice 
by introducing an extra spatial coordinate $x_5$. 
The DW fermion is formulated in the five-dimensional space-time, 
and its (five-dimensional) kernel is expressed as \cite{SDRS16}
\begin{eqnarray}
\hat K_5
&=&\hat {\Slash D}+m
+\frac{r}{2a}\sum_{\mu=\pm 1}^{\pm 4} \gamma_\mu (\hat U_\mu-1)
+\gamma_5\hat \partial_5+M(x_5) \cr
&=&\frac{1}{2a}\sum_{\mu=1}^{4} \gamma_\mu (\hat U_\mu-\hat U_{-\mu})+m
+\frac{r}{2a}\sum_{\mu=1}^4 \gamma_\mu (\hat U_\mu+\hat U_{-\mu}-2)
+\gamma_5\hat \partial_5 \cr
&+&M(x_5),
\label{eq:DWFK}
\end{eqnarray}
where the last two terms are the kinetic and the mass terms in the fifth dimension. 
In this formalism, $x_5$-dependent mass $M(x_5)$ is introduced 
as shown in Fig.\ref{fig:DWF}, where $M_0=|M(x_5)|=O(a^{-1})$ is taken to be large. 
Since $\hat K_5$ includes only kinetic and mass terms on the extra coordinate $x_5$, 
the eigenvalue problem is easily solved in the fifth direction. 
In this construction, there appear chiral zero modes \cite{K92,S93FS95}, i.e., 
a left-handed zero mode localized around $x_5=0$ and 
a right-handed zero mode localized around $x_5=N_5$, as shown in Fig.\ref{fig:DWF}. 
The extra degrees of freedom in the fifth dimension can be integrated out 
in the generating functional, with imposing the Pauli-Villars regularization
to remove the UV divergence.

For the five-dimensional DW fermion kernel $\hat K_5$, 
we define its eigenmode $|\nu \rangle$ and eigenvalue $\Lambda_\nu$ as 
\begin{eqnarray}
\hat K_5|\nu \rangle =i \Lambda_\nu |\nu \rangle, \quad 
\Lambda_\nu \in {\bf C}, \quad \qquad
\sum_\nu |\nu \rangle \langle \nu|=1+O(a).
\label{eq:QCR3}
\end{eqnarray}
Note that each term of $\hat K_5$ in Eq.(\ref{eq:DWFK}) 
connects only the neighboring site or acts on the same site 
in the five-dimensional space-time, 
and we can use almost the same technique as the Wilson fermion case. 
We consider the functional trace on a lattice with $N_t =4l+1$, 
\begin{eqnarray}
J \equiv {\rm Tr}(\hat U_4^{2l+1} \hat{K}_5^{2l}) 
\simeq \sum_\nu \langle \nu|\hat{U}_4^{2l+1}{\hat{K}_5}^{2l}|\nu \rangle
=\sum_\nu(i \Lambda_\nu)^{2l} \langle \nu|\hat{U}_4^{2l+1}| \nu \rangle,
\end{eqnarray}
where the quasi-completeness for $|\nu \rangle$ in Eq.(\ref{eq:QCR3}) 
is used. 
$J \equiv {\rm Tr}(\hat U_4^{2l+1} \hat K_5^{2l})$ 
is expressed as a sum of 
products of $\hat U$ with other factors, 
and each product gives a trajectory as shown in Fig.\ref{fig:FK} 
in the projected four-dimensional space-time.
Among the trajectories, 
only the Polyakov loop $L_P$ can form a closed loop 
and survives in $J$, i.e., $J \propto L_P$. 
Thus, apart from an $O(a)$ error, one finds \cite{SDRS16}
\begin{eqnarray}
L_P \propto
\sum_\nu \Lambda_\nu^{2l} \langle \nu |\hat{U}_4^{2l+1}| \nu \rangle. 
\label{eq:DWPL}
\end{eqnarray} 

After integrating out the extra degrees of freedom in the fifth dimension 
in the generating functional, one obtains 
the four-dimensional physical-fermion kernel $\hat K_4$ \cite{S93FS95}. 
The physical fermion mode is given by 
the eigenmode $|n \rangle \rangle$ of $\hat K_4$ with its eigenvalue $\tilde \lambda_n$, 
\begin{eqnarray}
\hat K_4 | n \rangle \rangle =i \tilde \lambda_n |n \rangle \rangle, \quad 
\tilde \lambda_n \in {\bf C}.
\end{eqnarray}
We find that the four-dimensional physical fermion eigenvalue 
$\tilde \lambda_n$ of $\hat K_4$
is approximatly expressed with the eigenvalue 
$\Lambda_\nu$ of the five-dimensional DW kernel $\hat K_5$ as \cite{SDRS16}
\begin{eqnarray}
\Lambda_\nu = \tilde \lambda_{n_\nu} +O(M_0^{-2})
=\tilde \lambda_{n_\nu} +O(a^2). 
\end{eqnarray}

Combining with Eq.(\ref{eq:DWPL}), apart from an $O(a)$ error, 
we obtain \cite{SDRS16}
\begin{eqnarray}
L_P \propto
\sum_\nu \tilde \lambda_{n_\nu}^{2l} 
\langle \langle \nu |\hat{U}_4^{2l+1}| \nu \rangle \rangle, 
\end{eqnarray} 
and find small contribution from low-lying physical-fermion modes of $\hat K_4$ 
to the Polyakov loop $L_P$, 
due to the suppression factor $\tilde \lambda_{n_\nu}^{2l}$ in the sum.

\section{The Wilson loop and Dirac modes on arbitrary square lattices}

\begin{figure}[h]
\begin{center}
\includegraphics[scale=0.5]{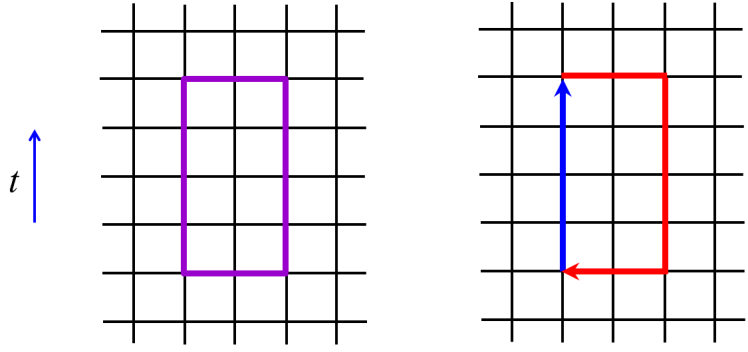}
\caption{(a) The Wilson loop $W$ on a $R \times T$ rectangle. 
(b) The factorization of the Wilson-loop operator as a product of 
$\hat U_{\rm staple}\equiv \hat U_k^R \hat U_{-4}^T \hat U_{-k}^R$ 
and $\hat U_4^T$ \cite{SDI16}. Here, $R$, $T$ and the lattice size 
$N_s^3 \times N_t$ are arbitrary.
}
\label{fig:WL}
\end{center}
\end{figure}

In the following, we investigate the role of low-lying Dirac modes to 
the Wilson loop $W$, the inter-quark potential $V(R)$ and 
the string tension $\sigma$ (quark confining force), 
on {\it arbitrary} square lattices with any number of $N_t$ \cite{SDI16}.
We note that the ordinary Wilson loop of the $R \times T$ rectangle 
on the $t$-$x_k$ ($k=1,2,3$) plane is expressed by the functional trace, 
\begin{eqnarray}
W \equiv {\rm Tr}_{c} \hat U_k^R \hat U_{-4}^T \hat U_{-k}^R \hat U_4^T
={\rm Tr}_{c} \hat U_{\rm staple} \hat U_4^T,
\end{eqnarray}
where the ``staple operator'' $\hat U_{\rm staple}$ is defined by 
\begin{eqnarray}
\hat U_{\rm staple} \equiv \hat U_k^R \hat U_{-4}^T \hat U_{-k}^R.
\end{eqnarray}
In fact, the Wilson-loop operator is factorized to be 
a product of $\hat U_{\rm staple}$ and $\hat U_4^T$, 
as shown in Fig.\ref{fig:WL}.

\subsection{Even $T$ case}

In the case of even number $T$, we consider the functional trace,
\begin{eqnarray}
J \equiv {\rm Tr}_{c,\gamma} (\hat U_{\rm staple} \hat {\Slash D}^T)
=\sum_{n} \langle n| \hat U_{\rm staple} \hat {\Slash D}^T |n \rangle 
= (-)^{\frac{T}{2}}
\sum_{n} \lambda_n^T \langle n| \hat U_{\rm staple} |n \rangle,
\end{eqnarray}
where we use the completeness of the Dirac mode, 
$\sum_n |n\rangle \langle n|=1$.
With a parallel argument in Sec.~3, one finds at each lattice gauge configuration 
\begin{eqnarray}
J &=& \frac{1}{2^T} {\rm Tr}_{c,\gamma} [ \hat U_{\rm staple} 
     \{ \sum_{\mu=1}^{4} \gamma_\mu (\hat U_\mu-\hat U_{-\mu}) \}^T ]
 = \frac{1}{2^T} {\rm Tr}_{c,\gamma} [\hat U_{\rm staple} (\gamma_4 \hat U_4)^T]
\cr
 &=& \frac{4}{2^T} {\rm Tr}_{c} (\hat U_{\rm staple} \hat U_4^T) 
 = \frac{4}{2^T} W,
\end{eqnarray}
where, to form a loop in the functional trace, 
$\hat U_4$ has to be selected in all the 
$\hat{\Slash D} \propto \sum_\mu \gamma_\mu (\hat U_\mu-\hat U_{-\mu})$ 
in $\hat{\Slash D}^T$. 
All other terms correspond to non-closed lines and give exactly zero, 
because of the definition of $\hat U_{\pm \mu}$ in Eq.(\ref{eq:LVO}).
We thus obtain \cite{SDI16}
\begin{eqnarray}
W = \frac {(-)^{\frac{T}{2}}2^T}{4}
\sum_{n} \lambda_n^T \langle n| \hat U_{\rm staple} |n \rangle,
\qquad
W \propto
\int_{-\infty}^\infty d\lambda \rho(\lambda)
\lambda^T U_{\rm staple}(\lambda), 
\end{eqnarray}
with $\rho(\lambda)\equiv \frac{1}{V}\sum_n \delta(\lambda-\lambda_n)$
and $U_{\rm staple}(\lambda_n) \equiv \langle n|\hat{U}_{\rm staple}| n \rangle$.
As in Sec.3, the reduction factor $\lambda^{T}$ 
cannot be cancelled by other factors, 
because of the finiteness of $\rho(0)$ and $U_{\rm staple}(\lambda)$ reflecting 
the Banks-Casher relation and the compactness of $U_{\rm staple} \in {\rm SU}(N_c)$.

Then, the inter-quark potential $V(R)$ is given by
\begin{eqnarray}
V(R) =-\lim_{T \to \infty} 
\frac{1}{T}{\rm ln} \langle W \rangle
= -\lim_{T \to \infty}\frac{1}{T}
{\rm ln} \left| \left\langle \sum_{n} 
(2 \lambda_n)^T \langle n| \hat U_{\rm staple} |n \rangle \right \rangle \right|, 
\end{eqnarray}
where $\langle \rangle$ denotes the gauge ensemble average. 
The string tension $\sigma$ is expressed by 
\begin{eqnarray}
\sigma =-\lim_{R,T \to \infty} \frac{1}{RT}{\rm ln} \langle W \rangle
= -\lim_{R,T \to \infty}\frac{1}{RT}
{\rm ln} \left | \left \langle \sum_{n} 
\lambda_n^T \langle n| \hat U_{\rm staple} |n \rangle \right \rangle \right|.
\label{eq:STeven}
\end{eqnarray}
Due to the reduction factor $\lambda_n^T$ in the sum, 
the string tension $\sigma$ or the quark confining force 
is unchanged by removing low-lying Dirac-mode contributions from Eq.(\ref{eq:STeven}).

\subsection{Odd T case }

In the case of odd number $T$, 
the similar results are obtained from 
\begin{eqnarray}
J &\equiv& {\rm Tr}_{c,\gamma} 
(\hat U_{\rm staple} \hat U_4 \hat{\Slash D}^{T-1})
=\sum_{n} \langle n| 
\hat U_{\rm staple} \hat U_4 \hat{\Slash D}^{T-1} |n \rangle 
\cr
&=& (-)^{\frac{T-1}{2}}
\sum_{n} \lambda_n^{T-1} \langle n| \hat U_{\rm staple} \hat U_4 |n \rangle.
\end{eqnarray}
Actually, one finds 
\begin{eqnarray}
J &=& \frac{1}{2^{T-1}} {\rm Tr}_{c,\gamma} [\hat U_{\rm staple} \hat U_4 
      \{\sum_{\mu=1}^{4} \gamma_\mu (\hat U_\mu-\hat U_{-\mu})\}^{T-1}]
\cr 
&=&\frac{1}{2^{T-1}}{\rm Tr}_{c,\gamma} [\hat U_{\rm staple} 
     \hat U_4 (\gamma_4 \hat U_4)^{T-1}] 
 = \frac{4}{2^{T-1}} {\rm Tr}_{c} \hat U_{\rm staple} \hat U_4^T 
 = \frac{4}{2^{T-1}} W,
\end{eqnarray}
and obtains for odd $T$ the similar formula \cite{SDI16}
\begin{eqnarray}
W &=& \frac {(-)^{\frac{T-1}{2}}2^{T-1}}{4}
\sum_{n} \lambda_n^{T-1} \langle n| \hat U_{\rm staple} \hat U_4|n \rangle, \\
W &\propto& 
\int_{-\infty}^\infty d\lambda \rho(\lambda)
\lambda^{T-1} U(\lambda), 
\label{eq:Wilsonodd}
\end{eqnarray}
with $U(\lambda_n) \equiv \langle n|\hat{U}_{\rm staple} \hat{U}_4| n \rangle$.

Then, the inter-quark potential $V(R)$ and the sting tension $\sigma$ 
are expressed as
\begin{eqnarray}
V(R) 
&=& -\lim_{T \to \infty}\frac{1}{T}
{\rm ln} \left| \left \langle \sum_{n} 
(2 \lambda_n)^{T-1} \langle n| \hat U_{\rm staple} \hat U_4|n \rangle 
\right \rangle\right|, 
\\
\sigma 
&=& -\lim_{R,T \to \infty}\frac{1}{RT}
{\rm ln} \left | \left \langle \sum_{n} 
\lambda_n^{T-1} \langle n| \hat U_{\rm staple} \hat U_4|n \rangle \right \rangle \right|.
\label{eq:STodd}
\end{eqnarray}
Owing to the reduction factor $\lambda_n^{T-1}$ in the sum, 
which cannot be cancelled by other factors, 
the string tension $\sigma$ is unchanged 
by the removal of low-lying Dirac-mode contributions from Eq.(\ref{eq:STodd}).

\section{Summary and conclusion}

In this paper, we have studied the relation between 
quark confinement and chiral symmetry breaking, 
which are the most important nonperturbative properties in low-energy QCD.
In lattice QCD formalism, 
we have derived analytical formulae on various ``confinement indicators'', 
such as the Polyakov loop, its fluctuations, 
the Wilson loop, the inter-quark potential and the string tension, 
in terms of the Dirac eigenmodes. 
We have also investigated the Polyakov loop in terms of the eigenmodes of 
the Wilson, the clover and the domain-wall fermion kernels, respectively,
and have derived similar formulae.

For all the relations obtained here, 
we have found that the low-lying Dirac modes contribution is 
negligibly small for the confinement quantities, 
while they are essential for chiral symmetry breaking.
This indicates no direct, one-to-one correspondence 
between confinement and chiral symmetry breaking in QCD.
In other words, there is some independence of quark confinement 
from chiral symmetry breaking.

This independence seems to be natural, 
because confinement is realized independently of quark masses 
and heavy quarks are also confined even without the chiral symmetry.
Also, such independence generally lead to different transition 
temperatures and densities for deconfinement and chiral restoration, 
which may provide a various structure of the QCD phase diagram.


\section*{Acknowledgments}
H.S. thanks A. Ohnishi for useful discussions. 
H.S. and T.M.D are supported in part by the Grants-in-Aid for Scientific Research 
(Grant No. 15K05076) and Grant-in-Aid for JSPS Fellows (Grant No. 15J02108) 
from Japan Society for the Promotion of Science.
K.R. and C.S. are partly supported by the Polish Science Center (NCN) 
under Maestro Grant No. DEC-2013/10/A/ST2/0010.
The lattice QCD calculations were done with SX8R and SX9 in Osaka University.

\end{document}